\documentclass[
    ,final            
  ]
  {aipproc}

\layoutstyle{6x9}

\def\nscat{N_{\rm \gamma e}}
\def\nel{n_{\rm e}}
\def\nt{n_{\rm t}}
\def\nuT{\nu_{\rm T}}
\def\nub{\nu_{\rm b}}
\def\ngamma{N_{\gamma}}
\def\mel{m_{\rm e}}
\def\fz{f_{\rm BH}}
\def\fznu{\fz (\nu)}
\def\Mbh{M_{\rm BH}}
\def\eps{\epsilon}
\def\ega{E_{\gamma}}

\def\Luv{L_{\rm UV}}
\def\fuv{f_{\rm UV}}
\def\rr{{\cal R}}
\def\Lbh{L_{\rm BH}}
\def\Lx{L_{\rm X}}
\def\fx{f_{\rm X}}
\def\sigkn{\sigma_{\rm KN}}
\def\sigknnu{\sigkn (\nu)}
\def\kb{k_{\rm B}}
\def\GC{\Gamma_{\rm C}}
\def\Gx{\Gamma_{\rm X}}
\def\Guv{\Gamma_{\rm UV}}
\def\TC{T_{\rm C}}
\def\Tuv{T_{\rm UV}}
\def\sigT{\sigma _{\rm T}}
\def\rcool{r_{\rm cool}}
\def\rc{r_{\rm c}}
\catcode`\@=11
\def\gsim{\ifmmode{\mathrel{\mathpalette\@versim>}}
    \else{$\mathrel{\mathpalette\@versim>}$}\fi}
\def\lsim{\ifmmode{\mathrel{\mathpalette\@versim<}}
    \else{$\mathrel{\mathpalette\@versim<}$}\fi}
\def\@versim#1#2{\lower 2.9truept \vbox{\baselineskip 0pt \lineskip
    0.5truept \ialign{$\m@th#1\hfil##\hfil$\crcr#2\crcr\sim\crcr}}}
\catcode`\@=12  

\begin{document}

\title{The role of Compton heating\\ in cluster cooling flows}

\author{L. Ciotti}{
  address={Astronomy Department, Bologna University, Italy}}
\author{J.P. Ostriker}{
  address={IoA, Cambridge, UK and Dept. of Astrophys. Sciences, 
                 Princeton University, NJ, USA}}
\author{S. Pellegrini}{
  address={Astronomy Department, Bologna University, Italy}
}

\begin{abstract}
Recent observations by Chandra and XMM-Newton demonstrate that the
central gas in "cooling flow" galaxy clusters has a mass cooling rate
that decreases rapidly with decreasing temperature. This contrasts the
predictions of a steady state cooling flow model. On the basis of
these observational results, the gas can be in a steady state only if
a steady temperature dependent heating mechanism is present;
alternatively the gas could be in an unsteady state, i.e., heated
intermittently. Intermittent heating can be produced by accretion on
the supermassive black hole residing in the central cluster galaxy,
via Compton heating. This mechanism can be effective provided that the
radiation temperature of the emitted spectrum is higher than the gas
temperature.  Here we explore whether this heating mechanism can be at
the origin of the enigmatic behavior of the hot gas in the central
regions of ``cooling flow'' clusters. Although several characteristics
of Compton heating appear attractive in this respect, we find that the
fraction of absorbed heating for realistic gas and radiation
temperatures falls short by two orders of magnitude of the
required heating.
\end{abstract}

\maketitle


\section{Introduction}

X-ray observations of the central regions of a large fraction of
galaxy clusters prior to $Chandra$ and $XMM-Newton$ were interpreted
in terms of cooling and condensing of the intracluster medium (ICM),
leading to a subsonic, steady central inflow called cooling flow
(Cowie \& Binney 1977, Fabian \& Nulsen 1977). The measurements of
radiative cooling times lower than a Hubble time within a cooling
radius ($\rcool$) of $\lsim $ few hundreds of kiloparsecs were at
the basis of this idea.  However, cool gas in the cluster cores (in
the form of massive-star formation or formation of low-mass stars,
optical emission line nebulae or cold gas) has never been discovered
in a quantity large enough to fit with the steady state cooling flow
predictions (e.g., Donahue \& Voit 2003). Moreover recent observations
from the $Chandra$ and $XMM-Newton$ satellites have ruled out the
simple steady state cooling flow model (e.g., Molendi \& Pizzolato
2001, Peterson et al. 2003).  In this model one expects emission from
gas over the entire temperature range (from $T_{\rm max}$, the ambient
ICM temperature, down to temperature values at which the ICM is
undetectable in the X-rays), {\it with the same mass cooling rate
$\dot M$ at each temperature}.  The strengths of a few observed
emission lines reveal how much gas cools through each temperature, and
high resolution spectroscopic observations now show a deficit of
emission relative to the cooling flow predictions from gas below 
$\sim T_{\rm max}/3$ (Peterson et al. 2003).  In addition, the spectra
show increasingly less emission at lower temperatures than the cooling
flow model would predict. Empirically, the differential luminosity
distribution $\Delta L/\Delta T\propto T^{1 \div 2}$, instead of being
temperature-independent as expected in the radiative cooling flow
model.  Lower spectral resolution observations, as from the $Chandra$
ACIS-S detector, also suggest significantly lower mass cooling rates
than obtained from previous analyses of $ROSAT$ and $ASCA$ data
(McNamara et al. 2000).

These observational results have produced an important astrophysical
puzzle and a number of ideas have been suggested to solve it, but none
has been proven conclusive yet. For example, cooling may be opposed by
heating (Soker et al. 2001) or thermal conduction may suppress cooling
(Ruskowkski \& Begelman 2002). `Cooling flows' generally have embedded
non-thermal radio sources and this association suggests that feedback
from an AGN may help suppress the cooling of the ICM. This feedback
may come from the impact of the radio jets on the cooling gas
(Reynolds et al. 2002, Omma et al. 2003) or from mixing and turbulent
heating as buoyant radio plasma rises through the ICM (Br\"uggen \&
Kaiser 2002). Another plausible (intermittent) heating source could be Compton
heating resulting from accretion of the cooling gas on the central
supermassive black hole of the central galaxy in the cluster core.
Although the gas over the body of the galaxy is optically thin,
numerical simulations showed that this mechanism is very effective in
heating the interstellar medium of giant elliptical galaxies (Ciotti
\& Ostriker 2001).  Here we explore whether it can provide effective
feedback also in the case of cluster cores.

The bolometric luminosity associated with accretion in the galactic
nucleus is $\Lbh =\eps \,\dot\Mbh \,c^2$, where $\dot\Mbh$ is the
accretion rate on the black hole, $\eps$ is the accretion efficiency
(usually spanning the range $0.001\lsim \eps \lsim 0.1$) and $c$ is
the speed of light. Taking $\dot\Mbh $ of the order of few 10s of
$M_{\odot}$ yr$^{-1}$, as suggested by the most recent estimates of
the mass accretion rate in `cooling flows' (Peterson et al. 2003), the
power available for heating of the ICM turns out to be $\sim 10^{47}$
erg s$^{-1}$.  The power required to balance the cooling of the ICM in
the `cooling flow' region must be of the order of its observed X-ray
luminosity; this goes from $\sim 10^{43}$ erg s$^{-1}$ in the Virgo
cluster up to $\sim 10^{45}$ erg s$^{-1}$ in the most massive clusters
like A1835. Therefore it seems that during the phases of accretion
there could be enough power to balance the cooling.  The problem is: How
much of $\Lbh$ is actually {\it trapped} by the inflowing gas and
therefore is effectively available for its heating? It is clear that
for extremely low opacities $\Lbh$ would be unable to affect the flow.
In the following we estimate how much of $\Lbh$ is absorbed by the ICM
in the core regions of galaxy clusters, for typical temperature and
density profiles.

\section{Setting the problem}

The number of photon--electron interactions per unit volume and in the
time interval $\Delta t$ at any radius $r$ in a plasma can be written as:
\begin{equation}
\nscat (\nu,r)=\ngamma (\nu,r)\times
{\sigkn(\nu) \nel (r)\over 4\pi r^2}={\Lbh (\nu,r)\Delta t\over h\nu}
\times{\sigkn (\nu)\nel (r)\over 4\pi r^2}, \label{inter}
\end{equation}
where $\nel (r)$ is the electron number density, $h$ is the Planck
constant and $\sigkn$ is the Klein-Nishina electron scattering
cross-section (Lang 1980):
\begin{equation}
\sigknnu = {3\sigT\over 4}\left\{ {1+x\over x^2}
\left [{2(1+x)\over 1+2x}-{\ln (1+2x)\over x}\right ]+{\ln (1 + 2x)\over 2x}-
{1+3x\over (1+2x)^2}\right\}, \label{klein}
\end{equation}
where $x\equiv\nu /\nuT$, $\nuT\equiv\mel c^2/h$ is the Thomson
frequency, $\mel$ is the electron mass and $\sigT$ is the Thomson
cross section. Here for simplicity we assume a {\it gray} absorption,
i.e.:
\begin{equation}
\Lbh(\nu,r)=\fznu\times \Lbh(r),\quad \int_0^{\infty}\fznu\,d\nu =1,
\label{gray}
\end{equation} 
where $\Lbh (r)$ is the bolometric accretion luminosity that
reaches the radius $r$ from the nucleus.

The gas Compton heating (or cooling) per unit frequency at radius $r$
is given by $\Delta E=-\nscat (\nu)\Delta\ega (\nu,T)$,
where $\Delta E$ is the internal energy per unit volume gained (or
lost) by the gas from radiation at frequency $\nu$, and
$\Delta\ega$ is the energy variation of a photon of frequency $\nu$
interacting with an electron of gas at temperature $T$.  A simple
approximation for the energy transfer factor is\footnote{This formula
reproduces the well known relations $\Delta\ega\sim 1.5\kb T-\mel c^2
x+O(1/x)$ for relativistic photon energy ($h\nu\gg \mel c^2$), and
$\Delta\ega\sim 4\kb Tx-\mel c^2x^2+O(x^3)$ in the classical limit
($h\nu \ll  \mel c^2$).}:
\begin{equation}
\Delta\ega(\nu,T)={x(1+3x^2/8)\over 1 + x^3}4\kb T-
                  {x^2(1+x^2)\over 1+x^3}\mel c^2 \label{transf}
\end{equation}
where $\kb$ is the Boltzmann constant.
After substitution of \eqref{inter} and \eqref{transf} in the
expression for the Compton heating, and integration over all
frequencies, one has:
\begin{equation}
{\Delta E\over \Delta t}
=-{\nel (r) \over\nt (r) } {E(r) \over 4\pi r^2}{\Lbh (r)\over \mel c^2}
{8\GC\over 3}\left [1-{\TC\over T(r)}\right ], \label{integ}
\end{equation}
where $\nt (r)$ is the total number density,
\begin{equation}
\GC \equiv \int _0^{\infty}
{(1+3x^2/8)\fz (\nu)\sigknnu\over 1 + x^3}\;d\nu \label{gc}
\end{equation}
and the spectral temperature $\TC$ is given by:
\begin{equation}
\TC\equiv{\mel c^2\over 4\kb\GC}\int _0^{\infty}{x(1+x^2)\fz (\nu)
\sigknnu\over 1+x^3}\; d\nu, \label{spect}
\end{equation}
where $\mel c^2/ 4\kb=1.48\times 10^9$ K.

Integrating on $r$ the equation of energy conservation $\partial
\Lbh(r)/\partial r= -4\pi r^2\partial E (r)/\partial t$, one has
\begin{equation}
\Lbh (r) = \Lbh (0) \exp \left\{
{8\over 3} { \GC \over \mel c^2} 
\int_0^r E (r)
{\nel (r)\over\nt (r)} \left( 1-{\TC\over T(r)}\right) dr \right\}. \label{lbh}
\end{equation}

From the above equation we can compute the amount of energy actually
trapped to heat the gas, for any gas temperature and density profile
and any spectral energy distribution of the nuclear photons.

We assume that the spectral energy distribution $\fznu $
is made of two
 distinct contributions. The first is a non thermal distribution of total
luminosity $\Lx(\nu)=\fx(\nu)\times \Lx$, with
\begin{equation}
\fx (\nu)={\xi_2\over\pi\nuT}\sin \left[{\pi(1-\xi_1)\over\xi_2}\right ]
\left ({\nub\over\nuT}\right )^{\xi_1+\xi_2-1}{x^{-\xi_1}\over
(\nub/\nuT)^{\xi_2}+x^{\xi_2}}, \label{fxnu}
\end{equation}
where $h\nub$ is the spectrum break energy and $\xi_1$ and
$\xi_1 + \xi_2$ are the spectral slopes at low and high frequencies. The second is
a blackbody distribution $\Luv(\nu)=\fuv(\nu)\times \Luv$ at a
temperature $\Tuv$, with
\begin{equation}
\fuv (\nu) ={15 h^4\over \pi^4\kb^4\Tuv^4}{\nu^3 \over\exp{(h\nu/\kb\Tuv)}-1}
           \simeq {8.17\times 10^{-43}\over \Tuv^4}
           {\nu^3\over\exp (h\nu/\kb\Tuv) -1}
           \,\,\,\, (s). \label{fuvnu}
\end{equation}

For these distributions $\int_0^{\infty}\fx(\nu)d\nu=1$ and
$\int_0^{\infty}\fuv(\nu)d\nu=1$.  We also assume that $\Luv\equiv \rr
\Lx$, where $\rr$ is a dimensionless parameter measuring the relative
importance of the ``UV bump'' with respect to the high energy part of
the spectral energy distribution.  Following this choice, the
frequency distribution of $\Lbh$ in \eqref{gray} can be written as
\begin{equation}
\fznu={\fx+\rr\fuv\over 1 +\rr}, \label{fznu}
\end{equation}
and $\Lbh (0)=\Lx+\Luv =(1+\rr)\Lx$.

\section{The results}

We first estimate the dependence of  $\GC$ and $\TC$, i.e., of
the integrals in \eqref{gc} and
\eqref{spect}, on the free parameters $\Tuv$, $\nub$, $\xi_1$,
$\xi_2$.  We adopt $\xi_1=0.9$ and $\xi_2=0.7$; these values reproduce
the observed spectral shapes of the X--ray and $\gamma$--ray emission
of AGNs (e.g., Nandra \& Pounds 1994; Lu \& Yu 1999).  The
coefficient $\GC$ is evaluated numerically by considering the range
$0.1\leq \nub/\nuT\leq 10$ for $\fx$ and the classical limit for
$\fznu$, i.e., $\kb\Tuv \ll h\nuT$. We obtain
\begin{equation}
\GC\equiv\Gx+\rr\Guv \simeq {0.77\times(\nub/\nuT)^{-0.06}+\rr\over 1+\rr}\times
                      \sigT. \label{uff}
\end{equation}
A similar evaluation of the dimensionless integral on the r.h.s. of 
\eqref{spect} gives 
\begin{equation}
\TC\simeq {8.6\times 10^7 \times(\nub/\nuT)^{0.196} +\rr\Tuv\over
           0.77\times (\nub/\nuT)^{-0.06} +\rr} \,\,\,\,\, (K). \label{tc}
\end{equation}
For observed values of $\nub/\nuT \sim 0.2$ and $\rr \sim 1$ (e.g.,
Fabian 1996), $\TC= 4\times 10^7$ K (independently of
$\Tuv$). This value for $\TC$ is very similar to that derived by
Sazonov et al. (2003) from composite mean QSOs spectra.

\begin{figure}
  \includegraphics[height=.6\textheight]{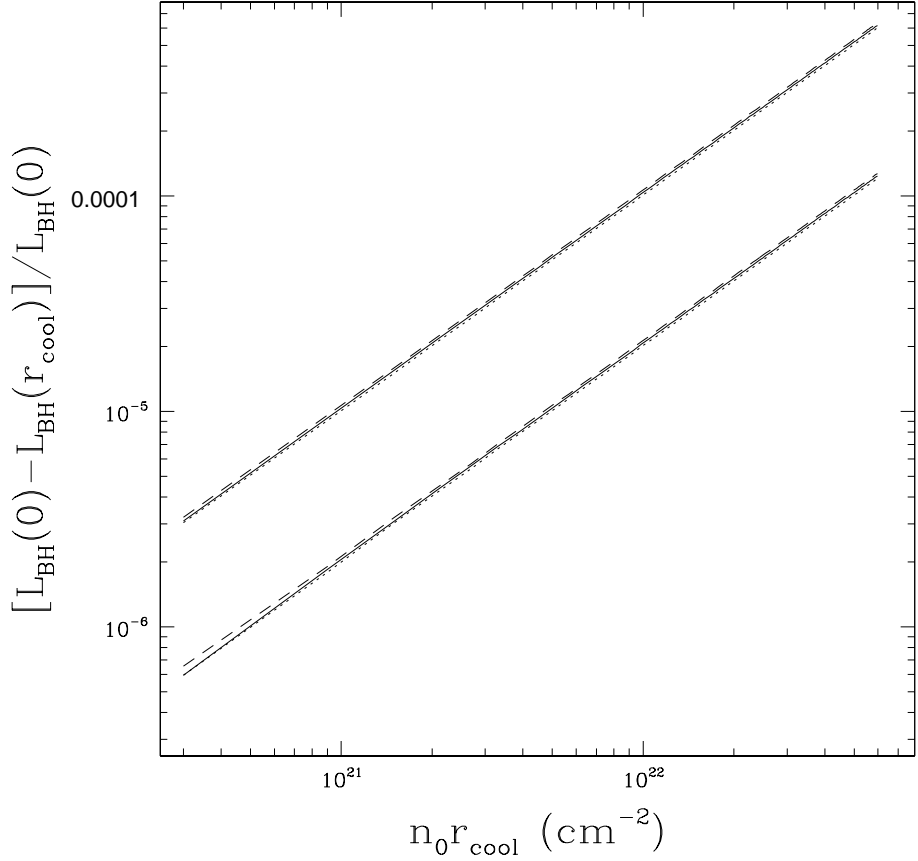} 
  \caption{The amount of nuclear radiation that actually heats the ICM
within $\rcool$. $\nub/\nuT$ is fixed at 0.2. The upper set of
curves refers to $\TC-T=5\times 10^7$ K, the lower set to $\TC-T=10^7$
K. Solid lines refer to $\rr=1$, dashed ones to $\rr=2$ and dotted
ones to $\rr=0.5$.}
\end{figure}

We next performed the integration in \eqref{lbh} to estimate $\Lbh
(r)- \Lbh (0)$, which is the amount of energy actually trapped by the
ICM within a radius $r$. The integration was extended out to the
`cooling radius' ($\rcool$), within which a `cooling flow' could
develop. For low $\TC $ values ($\sim 1-2$ keV, as allowed by the
study of Sazonov et al. 2003), $\rcool$ encloses the only region
within which the ICM temperature $T$ is lower than $\TC$. We assume
that the gas is isothermal at a temperature $T$ within $\rcool$.  The
adopted ICM density profile is a deprojection of the $\beta-$model
commonly used to reproduce the observed X-ray surface brightness
(i.e., Sarazin 1986): $\rho_{\rm
gas}=\rho_0[1+(r/\rc)^2]^{-3\beta/2}$, where $\rc$ is the core radius
($\sim 250$ kpc on average) and $\rho _0$ is the total particle
density at $r=0$. The integration in
\eqref{lbh} gives:
\begin{equation}
\Lbh (\rcool) = \Lbh (0) \exp \left[
-{4 \GC \over \mel c^2} {\nel \over \nt} \kb (\TC-T) n_0\rc f\left(
{\rcool\over\rc}, \beta \right) \right] \label{lbhint}
\end{equation}
where $n_0=\rho _0/\mu m_p$ with $\mu m_p$ the average particle mass.
For our choice of $\rcool\lsim\rc$, the function $f\approx
\rcool/\rc$ and is independent of $\beta$. This approximate value
for the integral is within 25\% of the true value, for
$\beta=0.3-0.9$.  In Fig. 1 we plot the amount of nuclear radiation
that actually heats the ICM within $\rcool$ as a function of $n_0
\rcool$. The range of variation for $n_0\rcool$ is chosen by
considering the observed ranges for $\rcool$ ($50-200$ kpc, Peterson
et al. 2003) and $n_0$ ($ 2\times 10^{-3}-0.1$ cm$^{-3}$).  Taking
$\nel /\nt =0.5$, the free parameters left in the expression for $\Lbh
(\rcool)$ in \eqref{lbhint} are $\nub/\nuT$, $\rr$ and $\TC-T$. In
Fig. 1 we consider a possible range for $\rr=0.5-2$ and two extreme
cases for the difference $\TC-T$, corresponding to a high $\TC$ value
[$\sim (5-6) \times 10^7$ K] and a low one ($\TC\sim 2\times 10^7$
K). Note that $\GC$ varies by $< 10$\% for $\nub/\nuT=0.1-1$, and so
the results are totally unaffected by variations of $\nub/\nuT$ in
this range.

\section{Conclusions}
From Fig. 1 it appears that the power actually available for heating
of the ICM within the cooling region goes from few$\times 10^{-6}
\Lbh(0)$, in the small clusters, up to $10^{-4}\Lbh (0)$ in the most
massive ones.  This makes Compton heating fall short by $\sim $two
orders of magnitude of the required heating and therefore an
unplausible mechanism to balance the cooling of the gas in the cluster
core, at variance with the situation in elliptical galaxies (Ciotti \&
Ostriker 2001).

\begin{theacknowledgments}
L.C. and S.P. thank the IoA for hospitality and financial support during a
visit where most of the results presented here were obtained.
\end{theacknowledgments}

\bibliographystyle{aipprocl} 


\begin{thebibliography}{}

\bibitem{1} Br\"uggen, M., Kaiser, C.R. 2002, Nature, 418, 301
\bibitem{2} Ciotti, L., Ostriker, J.P. 2001, Astrophys. J., 551, 131 
\bibitem{3} Cowie, L., Binney, J. 1977, Astrophys. J., 215, 723
\bibitem{4} Donahue, M., Voit, M. 2003, astro-ph/0308006
\bibitem{5} Fabian, A.C., Nulsen, P.E.J. 1977, Mon. Not. Roy. Astron. Soc., 
            180, 479
\bibitem{6} Fabian, A.C. 1996, Proc. 'Roentgenstrahlung from the Universe', 
            eds. Zimmermann, H.U., Truemper, J. and Yorke, H.; MPE Report 263,
            p. 403--408
\bibitem{7} Lang, K.R. 1980, Astrophysical Formulae, Springer-Verlag Berlin
\bibitem{8} Lu, Y., Yu, Q. 1999, Astrophys. J., 526, L5
\bibitem{9} McNamara, B., et al. 2000, Astrophys. J., 534, L135
\bibitem{10} Molendi, S., Pizzolato, F. 2001, Astrophys. J., 560, 194  
\bibitem{11} Nandra, K., Pounds, K.A. 1994, Mon. Not. Roy. Astron. Soc., 268,
             405
\bibitem{12} Omma, H., Binney, J., Bryan, G., Slyz, A. 2003, astro-ph/0307471
\bibitem{13} Peterson, J.R., et al. 2003, Astrophys. J., 590, 207
\bibitem{14} Reynolds, C.S., Heinz, S., Begelman, M.C. 2002, 
             Mon. Not. Roy. Astron. Soc., 332, 271
\bibitem{15} Ruskowski, M., Begelman, M.C. 2002, Astrophys. J., 581, 223
\bibitem{16} Sazonov, S.Y., Ostriker, J.P. \& Sunyaev, R.A. 2003
\bibitem{17} Soker, N., White, R.E.III, David, L.P., McNamara, B.R.
             2001, Astrophys. J., 549, 832

\end{thebibliography}

\end{document}